\documentclass[a4paper,11pt]{article}
\usepackage{jheppub} % for details on the use of the package, please see the JINST-author-manual
\usepackage{lineno}
\usepackage{subfigure}
\usepackage{slashed}
\newcommand{\bq}{{\bf q}}
\newcommand{\bp}{{\bf p}}
\newcommand{\bk}{{\bf k}}
\newcommand{\bea}{\begin{eqnarray}}
\newcommand{\eea}{\end{eqnarray}}

\title{\boldmath Solar Reflection of Inelastic Dark Matter}

\author[a,b,c]{Haipeng An}
\author[a]{Haoming Nie}
\affiliation[a]{Department of Physics, Tsinghua University, Beijing 100084, China}
\affiliation[b]{Center for High Energy Physics, Tsinghua University, Beijing 100084, China}
\affiliation[c]{Frontier Science Center for Quantum Information, Beijing 100084, China}  

\abstract{Solar-reflected dark matter (SRDM) consists of dark-matter particles up-scattered and accelerated by energetic electrons in the solar interior, producing a high-velocity tail that can enhance signals in direct-detection experiments, especially for MeV-scale masses. We consider an inelastic dark matter (iDM) model, in which solar scattering populates the excited state; subsequent de-excitation in terrestrial detectors releases the mass-splitting energy, substantially helping the energy release of the collision to be larger than the detector threshold. Using detailed Monte Carlo simulations, we generate the velocity and energy distributions of solar-reflected iDM over a range of dark-matter masses \(m_\chi\) and mass splittings \(\Delta\). We then compute event rates and energy depositions for current xenon and semiconductor experiments. Our results show that these experiments can place new constraints on the parameter space of MeV-scale iDM.
}

\begin{document}
\maketitle
\flushbottom

\section{Introduction}
\label{sec:intro}

Dark matter (DM), originally motivated by cosmological and astrophysical observations and now central to particle physics and cosmology, remains a profound mystery and a gateway to physics beyond the Standard Model. In recent years, theoretical developments have broadened the candidate mass range from \(10^{-22}\,\mathrm{eV}\) to hundreds of solar masses. Traditional weak-scale (WIMP-like) DM is now tightly constrained by experiments, shifting attention to light DM with MeV and sub-MeV masses. In this regime, direct detection relies on feeble DM–electron interactions that produce electron recoils in detector targets, and numerous searches have been performed for such signals~\cite{SENSEI:2023zdf, DAMIC-M:2023gxo, SuperCDMS:2023sql, XENON:2017vdw, XENON:2022ltv, LUX:2020car, LZ:2023poo, PandaX:2022xqx, CDEX:2022kcd, CDEX:2022dda, Gaior:2023wfr}. However, detecting electron recoils is challenging because the deposited energies are much smaller than those for nuclear recoils, leaving a substantial region of parameter space unexplored. On the experimental front, several ultra-low-threshold efforts leveraging novel sensor concepts and advanced readout electronics are being developed to probe these uncharted regions~\cite{Hochberg:2015pha,Oscura:2022vmi}.

We approach the problem from a complementary angle: boosting the initial dark-matter energy. A major advance in this direction is solar-reflected dark matter (SRDM)~\cite{An:2017ojc}. It was shown that a substantial population of DM can be reflected by the Sun before reaching Earth. In the Sun’s hot and dense plasma, a DM particle that scatters off a high-energy electron can be accelerated to keV energies, allowing it to exceed the thresholds of many current detectors. The energy transfer is most efficient when \(m_\chi \simeq m_e\), making the MeV mass range especially promising. Subsequent work has developed the SRDM framework for both contact interactions and massless mediators~\cite{An:2021qdl,Emken:2024nox}.

\begin{figure}
    \centering
    \includegraphics[width=0.7\linewidth]{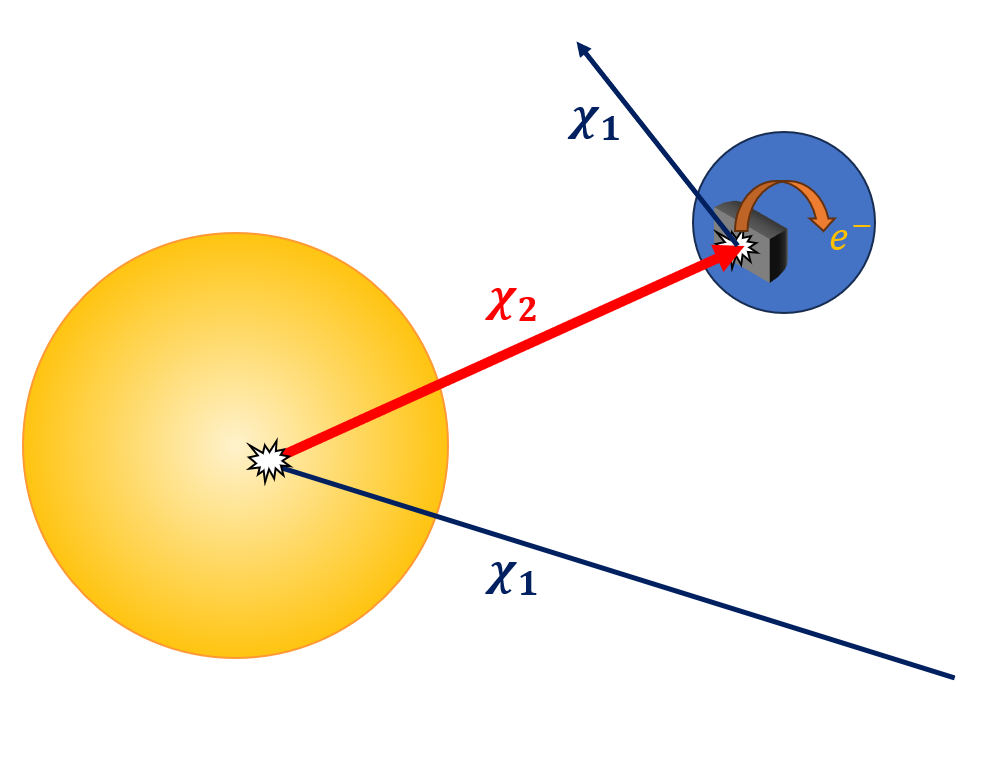}
    \caption{An illustration of solar-reflected inelastic DM.}
    \label{fig:inelastic_SRDM_illu}
\end{figure}

In this work, we study solar boosting in the context of inelastic dark matter (iDM). We consider a two-state DM system with a ground state \(\chi_1\) and an excited state \(\chi_2\), separated by a mass splitting \(\Delta \equiv m_2 - m_1 > 0\). We assume that the halo DM is entirely in the ground state \(\chi_1\) and that the DM–electron interactions are off-diagonal, so that collisions induce \(\chi_1 \leftrightarrow \chi_2\) transitions, while elastic \(\chi_1\)–electron scattering is neglected. A representative operator is \(\chi_1\chi_2 \,\bar e e\), with \(\chi_{1,2}\) taken as scalars; Majorana fermion DM is also possible. Because both solar boosting and direct detection occur in the non-relativistic regime, our results depend primarily on the interaction being inelastic, momentum independent, and spin independent. With these assumptions, up-scattering \(\chi_1 e \to \chi_2 e\) in DM detectors is kinematically forbidden unless the typical halo kinetic energy \(\sim \tfrac{1}{2} m_\chi v^2 \approx 10^{-6} m_\chi\) exceeds \(\Delta\). By contrast, if \(\chi_1\) enters the Sun, it can be up-scattered to \(\chi_2\) by hot solar electrons provided \(\Delta\) is below the electron kinetic energies. A fraction of the resulting \(\chi_2\) flux then reaches Earth and can de-excite to \(\chi_1\) inside a detector, releasing the splitting energy \(\Delta\) in addition to the recoil, often above threshold (see Fig.~\ref{fig:inelastic_SRDM_illu}). A brief single-scattering analysis was presented in Ref.~\cite{Baryakhtar:2020rwy}. Here we numerically simulate the \(\chi_2\) flux at Earth from solar reflection and derive constraints on the DM–electron cross section.

Inelastic DM models have been extensively studied~\cite{Tucker-Smith:2001myb,Tucker-Smith:2004mxa,Arina:2007tm,Chang:2008gd,Cui:2009xq,Lin:2010sb,DeSimone:2010tf,An:2011uq,Pospelov:2013nea,Giudice:2017zke,He:2024hkr}. Searches have been proposed and performed in direct-detection experiments such as XENON~\cite{Harigaya:2020ckz,Lee:2020wmh,Bramante:2020zos,Choi:2020ysq,An:2020tcg,Bell:2021zkr,Emken:2021vmf,Su:2023zgr} and SENSEI~\cite{Gu:2022vgb}, with additional proposals using heavy nuclei~\cite{Song:2021yar} and bubble chambers~\cite{PICO:2023uff}, as well as collider probes~\cite{Kang:2021oes,Li:2021rzt,Lu:2023cet}. Further avenues include cosmic-ray up-scattering~\cite{Bell:2021xff}, cosmic radiation~\cite{Finkbeiner:2007kk,Finkbeiner:2014sja}, and astrophysical signals~\cite{Alvarez:2023fjj}. The paper is organized as follows. In Section 2, we review SRDM for elastic interactions and outline our simulation framework. In Section 3, we compute the up- and down-scattering rates for iDM and present numerical results for solar-reflected iDM. In Section 4, we provide the direct-detection formalism for xenon and semiconductor targets and derive exclusion bounds from XENON and CDEX experiments.

\section{Overview of Solar Reflected Elastic DM}

\begin{figure}[htbp]
\centering
%\includegraphics[width=0.46\textwidth]{neplot.pdf}
%\qquad
%\includegraphics[width=0.46\textwidth]{Tplot.pdf}
\includegraphics[width=0.8\linewidth]{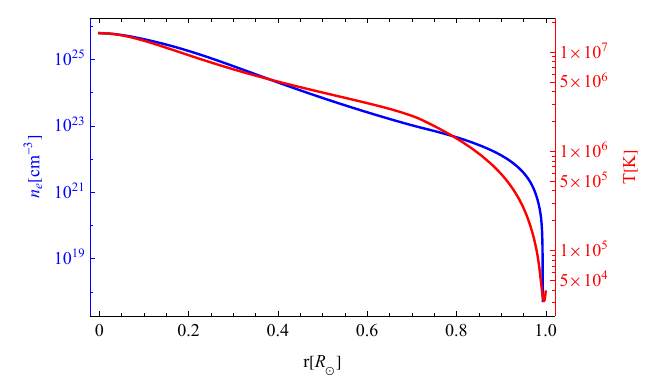}
\caption{These graphs show solar profile with respect to radius distance $r$ in unit of solar radius $R_\odot$. The blue line depicts electron number density, the red line depicts temperature.}
\label{solarprofile}
\end{figure}

The solar interior is a hot, dense plasma, with temperatures ranging from O(1 eV) at the surface to O(1 keV) in the core. In this work, we use the solar profile of Ref.~\cite{Bahcall:2004pz}, illustrated in Fig.~\ref{solarprofile}. The basic idea of solar-reflected DM is that when DM enters the Sun and scatters off high-energy electrons in the plasma, electrons can transfer energy to the DM and accelerate it. In the non-relativistic limit, the maximal DM kinetic energy after a single scatter is approximately
\[
E_\chi < E^{\text{max}}_{\chi} = \frac{4 E_e\, m_\chi m_e}{(m_\chi + m_e)^2},
\]
where \(E_e \sim T\) is the incident electron energy and \(m_\chi\) is the DM mass. The transfer is most efficient for \(m_\chi \approx m_e\).

To compute the reflected flux, we require the DM scattering rate on solar particles (electrons and ions). In general, the rate in a plasma can be written as
\begin{equation}
\Gamma_\chi = \frac{1}{2k_1^0}\!\int\!\frac{d^3 k_2}{2k_2^0(2\pi)^3}\!\int\!\frac{d^3 p_1}{2p_1^0(2\pi)^3} f(p_1)\!\int\!\frac{d^3 p_2}{2p_2^0(2\pi)^3}\,(2\pi)^4 \delta^{(4)}(k_1 + p_1 - k_2 - p_2)\!\sum_{\text{spins}}\!|\mathcal{M}|^2,
\label{generalGamma}
\end{equation}
where \(k_{1,2}\) are the incoming/outgoing DM four-momenta, \(p_{1,2}\) are the incoming/outgoing electron (or ion) four-momenta, and \(\sum_{\text{spins}}|\mathcal{M}|^2\) denotes the squared matrix element summed over the final and averaged over the initial spins. The phase-space occupation of solar particles at a given position is modeled by a Maxwell–Boltzmann distribution,
\[
f(p_1) = n \left(\frac{2\pi}{m T}\right)^{3/2} e^{-p_1^2/(2 m T)},
\]
with number density \(n\) and local temperature \(T\).

For elastic contact interactions between DM and electrons, Ref.~\cite{An:2021qdl} takes
\[
\sum_{\text{spins}}|\mathcal{M}|^2 = 16\pi (m_\chi + m_e)^2 \sigma_{\text{tot}},
\]
leading after straightforward manipulations to
\[
\Gamma_\chi\!=\!\frac{(m_\chi + m_e)^2 \pi \sigma_{\text{tot}} n_e m_e}{m_e^2 m_\chi^2}
\!\left(\!\frac{2\pi}{m_e T}\!\right)^2
\!\!\int\!\frac{d^3 q}{(2\pi)^3}\frac{1}{q}
\exp\!\left[-\frac{1}{2 m_e T}\!\left(\!\frac{m_e}{2 m_\chi}(2 k_1 \cos\theta_{\mathbf{k}_1\mathbf{q}}\!+\!q)\!+\!\frac{q}{2}\!\right)^{\!2}\right],
\]
and the corresponding differential rate
\begin{equation}
\frac{d\Gamma_\chi}{dq\, d\cos\theta_{\mathbf{k}_1\mathbf{q}}}
\simeq \mathcal{N}\, q\,
\exp\!\left[-\frac{1}{m_e T}\!\left(\frac{m_e}{2 m_\chi}(2 k_1 \cos\theta_{\mathbf{k}_1\mathbf{q}} + q) + \frac{q}{2}\right)^{\!2}\right],
\label{elastic_dGammadxqdcq}
\end{equation}
where \(k_1 \equiv |\mathbf{k}_1|\), \(q \equiv |\mathbf{q}|\), and \(\mathcal{N}\) is a normalization factor. For ion scattering, replace \(m_e\) with the ion mass. Since \(m_{\text{ion}} \gg m_\chi\), ion scattering primarily deflects the DM trajectory with little energy exchange; its impact was also studied in Ref.~\cite{An:2021qdl}.

The simulation strategy in Ref.~\cite{An:2021qdl} is the following:
\begin{itemize}
    \item Partition the solar interior into \(\sim 2000\) thin spherical shells with homogeneous properties. This allows the scattering probability within each shell to be computed without integrating along curved trajectories.
   
    \item Sample DM particles from infinity with speeds drawn from a Gaussian halo distribution \(\propto \exp(-v^2 / v_0^2)\) with \(v_0 \approx 220~\mathrm{km/s}\). Choose the impact parameter uniformly over a disk of radius \(4 R_\odot\) to account for gravitational focusing during propagation.
    
    \item For particles entering the Sun, compute the path length \(\Delta l\) in each shell. Taking the shells sufficiently thin, neglect gravitational variation within a shell. Use \(P_\chi = \Gamma_\chi \Delta l / v_\chi\) to stochastically determine whether a scatter (off an electron or ion) occurs in that shell. When a scatter occurs, generate the outgoing DM momentum according to Eq.~\eqref{elastic_dGammadxqdcq}. We work in the regime \(P_\chi \ll 1\), so at most one scatter per species is attempted per shell.
    
    \item Update the DM energy and angular momentum to identify the next shell and continue the propagation. If the particle exits the Sun with sufficient total energy to reach infinity, record its asymptotic kinetic energy.
    
    \item Accumulate a histogram of the outgoing particles to obtain a flux normalized to the impact disk area \(A_\rho = \pi (4 R_\odot)^2\), which we denote by \(F_{A_\rho}\). Note that each DM particle injected into the Sun represents a flux weighted by its initial velocity. Therefore, to obtain the correctly normalized reflected DM energy distribution, the histogram must be weighted by the initial velocity prior to normalization.

\end{itemize}

The SRDM flux at Earth is then
\[
\frac{d\Phi_{\text{SRDM}}}{dE_\chi} = \Phi_{\text{halo}} \times \frac{F_{A_\rho} A_\rho}{4\pi r_\oplus^2},
\]
where \(r_\oplus \simeq 1~\mathrm{AU}\). This flux only accounts for the DM particles that come close enough to the Sun so that they are reflected by solar particles or gravitationally affected by the Sun. The simulated flux in Ref.~\cite{An:2021qdl} was analyzed against XENON1T data~\cite{XENON1T}, yielding strong constraints around \(m_\chi \sim \mathrm{MeV}\).

\section{Scattering of Inelastic DM in the Sun}

Now we consider a generic two-state inelastic DM model with a contact interaction, where the masses of the two states are  
\(m_{\chi_1} = m_\chi - \frac{\Delta}{2}\) and \(m_{\chi_2} = m_\chi + \frac{\Delta}{2}\). For our purposes, the detailed model-building aspects are not important.

Including the mass splitting, the kinetic energy of an up-scattered DM particle that initially has negligible velocity and is struck by an electron with kinetic energy \(E_e\) is given by
\[
    E_\chi = \frac{\bigl(\sqrt{m_e m_\chi (E_e m_\chi - \Delta (m_e + m_\chi))} + m_\chi \sqrt{E_e m_e}\bigr)^2}{m_\chi (m_e + m_\chi)^2}.
\]
Evidently, the mass splitting consumes a significant fraction of the available energy transfer, thereby reducing the outgoing DM kinetic energy. Neglecting the initial halo DM kinetic energy, scattering is kinematically forbidden if the electron energy is smaller than the mass splitting, \(E_e < \Delta\). Even for \(E_e > \Delta\), scattering can remain forbidden for sufficiently small DM masses.

\begin{figure}
    \centering
    \includegraphics[width=0.5\linewidth]{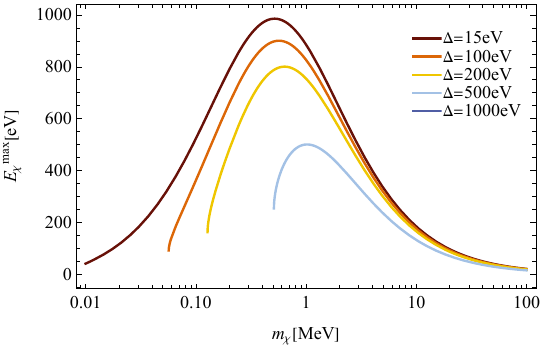}
    \caption{The relation between $E_\chi^{\text{max}}$ in electron-DM scattering in the Sun and the mass of DM. $E_e=1000\,\text{eV}$.}
    \label{fig:sun_scat}
\end{figure}

If the mass splitting is small $\Delta\ll m_\chi$, the DM-electron scattering amplitude of both up-scattering and down-scattering can be written as
\begin{equation}
    \sum_{\text{spin}}|\mathcal{M}|^2 = 16\pi(m_\chi+m_e)^2 \bar{\sigma}_{\text{tot}}.
\end{equation}
Now, let us compute the scattering rate of inelastic DM-electron(ion) scattering. The general formula is still the same as in Eq.~\eqref{generalGamma}:
\begin{equation}
    \Gamma_\chi^\eta = \frac{1}{2k_1^0}\int\frac{d^3 k_2}{2k_2^0(2\pi)^3}\int\frac{d^3 p_1}{2p_1^0(2\pi)^3}f(p_1)\int\frac{d^3 p_2}{2p_2^0(2\pi)^3}(2\pi)^4\delta^4(k_1+p_1-k_2-p_2)\sum_{\text{spin}}|\mathcal{M}|^2,
\end{equation}
where $k_1$ and $k_2$ are the four-momenta of the incoming and outgoing DM particles, $p_1$ and $p_2$ are the four-momenta of the incoming and outgoing electrons (or ions). If we define the momentum transfer $q=k_2-k_1$, and translate the integral variable $d^3 k_2\rightarrow d^3 q$, then we obtain:
\begin{equation}
    \Gamma_\chi^\eta = \frac{2\pi(m_\chi+m_e)^2\pi\bar{\sigma}_{\text{tot}}}{m_e^2 m_\chi^2}\!\int\!\frac{d^3 p_1}{(2\pi)^3}f_e(p_1)\!\int\!\frac{d^3 q}{(2\pi)^3}\delta\!\left(\!q^2\!\left(\!\frac{1}{2m_e}\!+\!\frac{1}{2m_\chi}\!\right)\!+\!\frac{\bk_1\cdot\bq}{m_\chi}\!-\!\frac{\bp_1\cdot\bq}{m_e}\!+\!\eta\Delta\!\right),
\end{equation}
where $\eta = +1$ for up-scattering and $\eta = -1$ for down-scattering. Following the same procedure in Ref.~\cite{An:2021qdl} and integrating out the angular part of $\bp_1$ first, we arrive at
\begin{equation}
    \Gamma_\chi^\eta = \frac{2\pi(m_\chi+m_e)^2\pi\bar{\sigma}_{\text{tot}}}{m_e^2 m_\chi^2}\int\frac{d^3 q}{(2\pi)^3}\int_{p_{1\text{min}}}\frac{p_1^2 dp_1}{(2\pi)^2}f_e(p_1)\frac{m_e}{p_1 q},
\end{equation}
where
\begin{equation}
    p_{1\text{min}}=\left|q\left(\frac{m_e}{2m_\chi}+\frac{1}{2}\right)+\frac{m_e}{m_\chi}\frac{\bk_1\cdot\bq}{q}+\frac{m_e}{q}\eta\Delta\right|.
\end{equation}
To simplify the expression in further integrals, we non-dimensionalize the variables as $x_\chi = v_\chi\sqrt{m_e/T}$, $x_q=q/\sqrt{m_e T}$, $x_{k_1}=k_1/\sqrt{m_e T}$, $\delta = \Delta/\sqrt{m_e T}$, $c_q = \cos\theta_{\bk_1 \bq}$, therefore, 
\begin{equation}
    \begin{split}
    \Gamma_\chi^\eta = &\frac{(m_\chi+m_e)^2\pi\bar{\sigma}_{\text{tot}}}{m_e^2 m_\chi^2}\frac{n_e m_e\sqrt{m_e T}}{(2\pi)^{3/2}}\int x_q d x_q d c_q n_e \\
    &\exp\left[-\frac{1}{2}\left(x_q\left(\frac{m_e}{2m_\chi}+\frac{1}{2}\right)+\frac{m_e}{m_\chi}x_{k_1}c_q+\eta\sqrt{\frac{m_e}{T}}\frac{\delta}{x_q}\right)^2\right].
    \end{split}
\end{equation}
The differential rate can then be expressed as
\begin{equation}
    \begin{split}
    \frac{d\Gamma_\chi^\eta}{dx_q dc_q}=&\frac{(m_\chi+m_e)^2\pi\bar{\sigma}_{\text{tot}}}{m_e^2 m_\chi^2}\frac{n_e m_e\sqrt{m_e T}}{(2\pi)^{3/2}}x_q\\
    &\exp\left[-\frac{1}{2}\left(x_q\left(\frac{m_e}{2m_\chi}+\frac{1}{2}\right)+\frac{m_e}{m_\chi}x_{k_1}c_q+\eta\sqrt{\frac{m_e}{T}}\frac{\delta}{x_q}\right)^2\right].\label{inelastic_dGammadxqdcq}
    \end{split}
\end{equation}
If we further integrate out $c_q$, then
\begin{equation}
    \begin{split}
    \Gamma_\chi^\eta =& \frac{(m_\chi+m_e)^2\bar{\sigma}_{\text{tot}}}{m_e^2 m_\chi^2}\frac{n_e m_\chi}{4\pi}\frac{\sqrt{m_e T}}{x_{k_1}}\!\int\! x_q d x_q \!\left\{\!\text{erf}\!\left[\!\frac{1}{\sqrt{2}}\!\left(\!\frac{m_e}{m_\chi}x_{k_1}\!+\!\left(\!\frac{m_e}{2m_\chi}+\frac{1}{2}\!\right)x_q\!+\!\eta\sqrt{\frac{m_e}{T}}\frac{\delta}{x_q}\!\right)\!\right]\right.\\
    &-\left.\text{erf}\left[\frac{1}{\sqrt{2}}\left(-\frac{m_e}{m_\chi}x_{k_1}+\left(\frac{m_e}{2m_\chi}+\frac{1}{2}\right)x_q+\eta\sqrt{\frac{m_e}{T}}\frac{\delta}{x_q}\right)\right]\right\}.
    \end{split}
\end{equation}

As a result, the probability that a DM particle is scattered by an electron along an infinitesimal distance $\Delta l$ is
\begin{equation}
    P^\eta = \Gamma_\chi^\eta\frac{\Delta l}{v_\chi}. \label{P_inelastic}
\end{equation}
The scatterings between DM particles and ions follow the same formula, with mass and density replaced by $m_{\text{ion}}$ and $n_{\text{ion}}$. In this paper, we assume that the cross section between a single nucleon and DM is the same as the electron-DM cross section, and the ion cross section is $\bar{\sigma}_{\text{ion}}=A^2\bar{\sigma}_e$, where $A$ is the number of nucleons in a nucleus. In the solar interior, only the $^1\text{H}$ and $^4\text{He}$ ions are sufficiently abundant to be considered.

\begin{figure}
    \centering
    \includegraphics[width=0.7\linewidth]{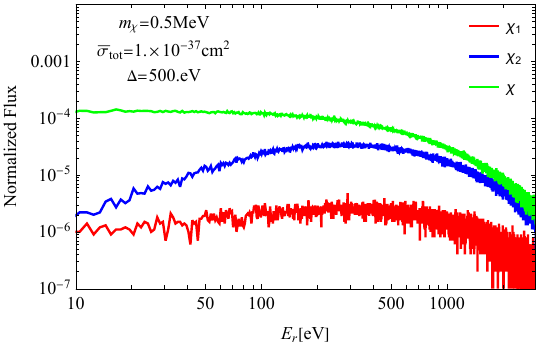}
    \caption{Normalized solar reflected inelastic DM spectrum. $E_r$ stands for reflected DM kinetic energy. $\chi_1$ stands for ground state DM, $\chi_2$ stands for excited DM. $\chi$ stands for elastic DM with same mass and characteristic cross section.}
    \label{fig:inelastic}
\end{figure}

In simulations, the scattering events are determined by comparing randomly generated numbers in the interval $[0,1]$ to $1-e^{-P^\eta}$ from Eq.~\eqref{P_inelastic}. The outgoing momentum and energy are randomly generated according to the distribution corresponding to Eq.~\eqref{inelastic_dGammadxqdcq}. After scattering, the DM particle transits from the ground state to the excited state or vice versa. The resulting flux on Earth contains both states, while only DM in the excited state is beneficial for sensitivity because it enhances energy deposit. A benchmark result is shown in Fig.~\ref{fig:inelastic}. The excited state DM flux is much larger than ground state DM flux in the high-energy tail, because one-scattering dominates at this cross section. In the elastic case in Ref.~\cite{An:2021qdl}, the ion scatterings result in little change in the DM energy, but they can bend the trajectory and increase the time of DM staying in the Sun, enabling more electron scattering possibilities. In the inelastic case, additionally, ion scatterings can alter the state of the DM particle and increase the energy deposit corresponding to the mass splitting.

If the cross section becomes larger, multiple scattering becomes common. Henceforth, DM in the excited state and ground state approach equally populated, especially for the high-energy tail $E_r>1\,\text{keV}$. For sufficiently large cross sections, the two states should approach thermalization. Therefore, the ground state will be much more populated than the excited states, especially for large mass splitting, but this situation is not reached in our parameter region of interest.

\begin{figure}
    \centering
    %\subfigure[\label{fig:a}]{\includegraphics[width=0.7\linewidth]{inelastic_plot_0.pdf}}\\
    %\subfigure[\label{fig:b}]{\includegraphics[width=0.7\linewidth]{inelastic_plot_10.pdf}}\\
    %\subfigure[\label{fig:c}]{\includegraphics[width=0.7\linewidth]{inelastic_plot_30.pdf}}
    \includegraphics[width=0.98\linewidth]{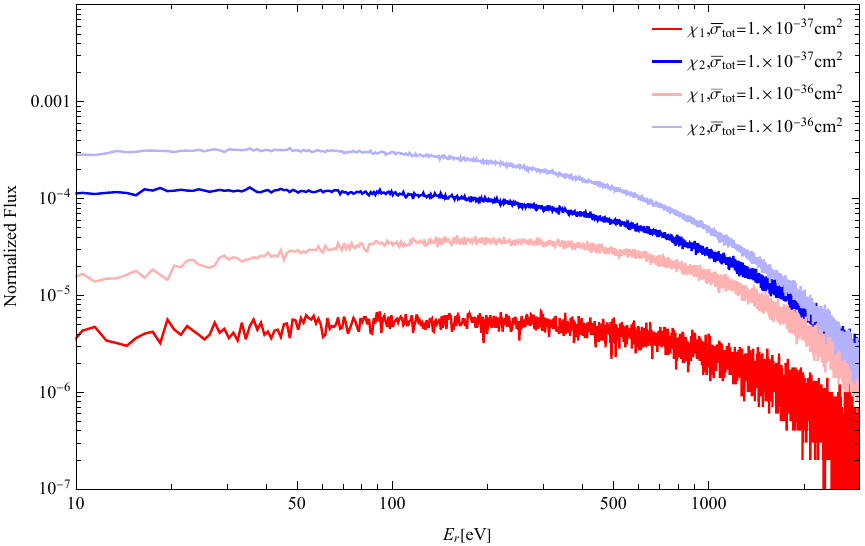}
    \caption{Normalized solar reflected inelastic DM spectrum for different $\bar{\sigma}_{\text{tot}}$. $E_r$ stands for reflected DM energy. $\chi_1$ stands for ground state DM, $\chi_2$ stands for excited DM. All the lines have $m_\chi=0.5\,\text{MeV}$.}
    \label{fig:enter-label}
\end{figure}

\section{Direct Detection of Solar Reflected Inelastic DM}
In this section, we use the reflected flux to set constraints in parameter space by the direct-detection experiment signal, focusing on XENONnT S1+S2 data~\cite{XENON:2022ltv}, XENON1T S2only data~\cite{XENON:2019gfn} and CDEX-10~\cite{CDEX:2022kcd} data. In the XENONnT experiment, if both S1 and S2 signals are considered, the recoil energy should be larger than about 1 keV, in contrast to the 160 eV threshold of CDEX-10. Therefore, the relation between exclusion limits and mass splittings can be clearly discussed. If we only consider the S2 signal, the detection threshold can be significantly reduced. Therefore, we also show the exclusion limits by XENON1T S2only data for comparison.

\begin{figure}
    \centering
    \subfigure[$E_\chi=1000\,\text{eV}$]{
        \includegraphics[width=0.47\linewidth]{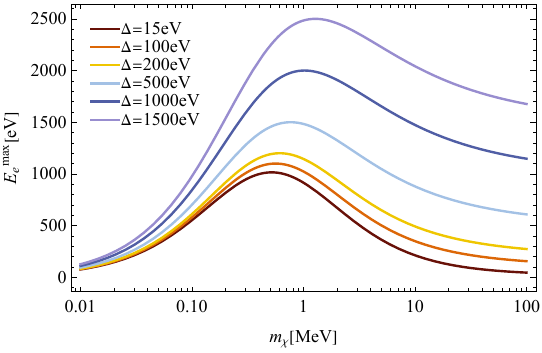}
        \label{fig:detector_scat_a}
    }
    \subfigure[$E_\chi=10\,\text{eV}$]{
        \includegraphics[width=0.47\linewidth]{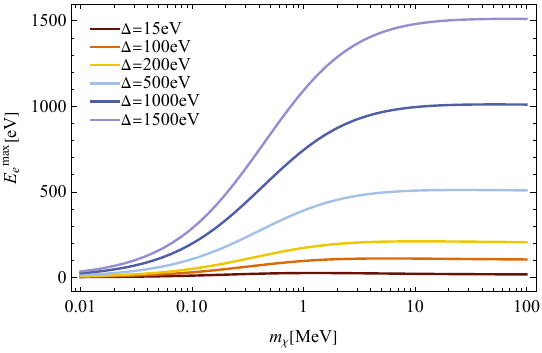}
        \label{fig:detector_scat_a2}
    }
    \subfigure[$\Delta=200\,\text{eV}$]{
        \includegraphics[width=0.47\linewidth]{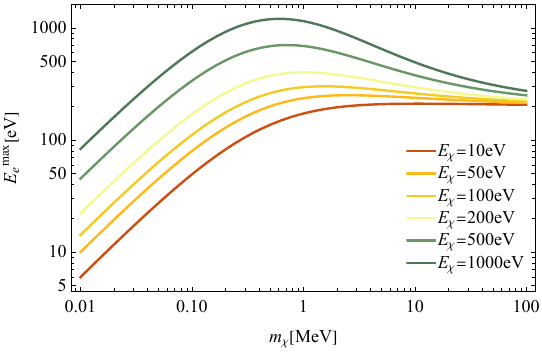}
        \label{fig:detector_scat_b}
    }
    \caption{The relation between $E_e^{\text{max}}$ in DM-electron scattering in the detector and the mass of DM.}
    \label{fig:detector_scat}
\end{figure}

In the detector target, an excited DM particle ionizes electrons through down-scatterings in the material and releases energy corresponding to the mass splitting. In general, in a 2-to-2 inelastic scattering event, the released energy from mass splitting will be mostly transferred to the particle with smaller mass. If we approximate the initial electron as a static free electron, the recoil energy of the electron can be estimated as
\begin{equation}
    E_e<E^{\text{max}}_e=\frac{(\sqrt{m_e m_\chi (E_\chi m_e+\Delta (m_e+m_\chi))}+m_e \sqrt{E_\chi m_\chi})^2}{m_e (m_e+m_\chi)^2}.
\end{equation}
From the form of the expression we can deduce that when $m_\chi<m_e$, the recoil energy is not sensitive to the mass splitting. After $m_\chi>m_e$, the larger the DM mass is, the more sensitive the recoil energy is to mass splitting. This observation is confirmed by the plot in Fig.~\ref{fig:detector_scat_a}. Therefore, the detection of inelastic DM can be divided into two situations. When $m_\chi<m_e$, the signal is sensitive to the kinetic energy of SRDM gained from DM-electron scattering and the mass splitting is not important, as shown in Fig.~\ref{fig:detector_scat_b}. On the other hand, when $m_\chi>m_e$, the kinetic energy is irrelevant and the mass splitting dominates the energy deposit. Accounting for the fact that low-mass DM scattering in the Sun is suppressed by large mass splittings mentioned in Section 3, we expect that large mass splittings enhance the signal for large-mass DM and suppress the signal for small-mass DM.

For the contact interaction case, the DM-electron interaction strength is conventionally represented by a reference cross section. In the iDM case, if the mass splitting $\Delta\ll m_\chi$, then the cross section is approximately the same for both up-scatterings and down-scatterings. Therefore, it is still possible to represent the interaction by a single cross section:
\begin{equation}
    %\bar{\sigma}_{\text{tot}}=\frac{16\pi\alpha\alpha_D\mu_e^2}{m_V^4},
    \bar{\sigma}_{\text{tot}}=\frac{4g^2m_e^2}{16\pi(m_\chi+m_e)^2},
\end{equation}
%where $m_V$ is the mediating dark photon mass, and $\alpha_D=e_D^2/4\pi$, where $e_D$ is the dark gauge coupling of DM.
where $g$ is the effective coupling of a 4-particle vertex $O_4=\chi_1\chi_2\bar{\psi}\psi$, where $\chi_1$ and $\chi_2$ are effective scalar fields, and $\psi$ is the electron field.

The general form of the velocity-averaged electron recoil cross section with respect to the electron recoil energy $E_e$ can be written as
\begin{equation}
    \frac{d\langle\sigma_e\rangle}{d\ln E_e}=\frac{\bar{\sigma}_{\text{tot}}}{8\mu_e^2}\int dq\left\{q|f_e(p_e,q)|^2\left[\eta(E_{\text{min}}^+(q,\Delta E))+\eta(E_{\text{min}}^-(q,\Delta E))\right]\right\}.
\end{equation}
In this formula, $|f_e(p_e,q)|^2$ is the electron form factor. Depending on the material and initial electron state, it can be either an atomic bound state ionization form factor or a valence shell Bloch state ionization form factor. It is a function of the momentum transfer $q$ and the final electron momentum $p_e$. The recoil energy is $E_e=p_e^2/2m_e$. The function $\eta$ represents an averaged squared inverse velocity, accounting for both up-scattering $E_{\text{min}}^+$ and down-scattering $E_{\text{min}}^-$:
\begin{equation}
    \eta(E_{\text{min}}^\pm)=\int_{E_{\text{min}}^\pm}dE_\chi\frac{m_\chi}{2E_\chi}\frac{1}{\Phi_{\text{halo}}}\frac{d\Phi_{SR}}{dE_\chi},
\end{equation}
where
\begin{equation}
    E^{\pm}_{\text{min}}=\frac{1}{2}m_\chi (v_{\text{min}}^\pm)^2=\frac{1}{2}m_e\left|\text{max}\left(\frac{\Delta E}{q}+\frac{q}{2m_\chi}\pm\frac{\Delta}{q},0\right)\right|^2.
\end{equation}
The total ionization rate is then given by
\begin{equation}
    \frac{dR_e}{d\ln E_e}=N_T\Phi_{\text{halo}}\frac{d\langle\sigma_e\rangle}{d\ln E_e},
\end{equation}
where $N_T$ is the number of atoms in the target.

In the XENON experiment, electron recoil induces an ionization signal in xenon (S2). And for $E_e\gtrsim 1\,\text{keVee}$, it also induces a scintillation signal (S1). In this paper, XENONnT S1+S2 data from Ref.~\cite{XENON:2022ltv} is used. For xenon, we only have the atomic ionization form factors $|f_{nl}(p_e,q)|^2$ corresponding to the atomic orbit $nl$. The calculations of the form factors $|f_{nl}(p_e,q)|^2$ follow the instructions of Ref.~\cite{Essig:2017kqs}, and the atomic wave functions are taken from Ref.~\cite{Bunge1993RHF}. We divide the energy range 0--30keV into 10 bins, the expected event counts corresponding to exposure time $T$ can be written as
\begin{equation}
    N_i=T\int_{\ln E_{e,\text{min},i}}^{\ln E_{e,\text{max},i}} d\ln E_e\,\text{Eff}(E_e)\frac{dR_e}{d\ln E_e},
\end{equation}
where $i=1,\dots,10$. Here, $\text{Eff}(E_e)$ is the efficiency function taken from Ref.~\cite{XENON:2022ltv}, $E_{e,\text{min},i}$, and $E_{e,\text{max},i}$ are the lower and upper bounds of each energy bin. With $N_i$, accompanied by the measured event counts $S_i$ and the background model counts $B_i$ (also from Ref.~\cite{XENON:2022ltv}) in each bin, we can construct a likelihood ratio statistic:
\begin{equation}
    -2\ln\lambda=\sum_{i=1}^{10}\left[-\ln\left(\frac{S_i}{N_i+B_i}\right)+\frac{(S_i-N_i-B_i)^2}{N_i+B_i}\right].
\end{equation}
For a sufficiently large sample, this statistic approaches a $\chi^2(10)$-distributed random variable. Therefore, if $-2\ln\lambda>18.307$, the hypothesis of the DM model is rejected at a significance level 0.05, therefore this parameter point is excluded.

For the XENON1T experiment, we adapt the publicly-available S2only data and analysis method from Ref.~\cite{XENON:2019gfn}. The region of interest (ROI) of photoelectron (PE) is taken to be the recommended value (165.3, 271.7). The calculation of expected event number is similar to XENONnT S1+S2.

The target material of the CDEX experiments is single-crystal semiconductor germanium. Therefore, the total event rate is the sum of Bragg scattering rate of inner shell bound states and valence shell Bloch state scattering rate. The calculations of the form factors follow the previous paper~\cite{An:2025cik}. For inner shell electrons, the scattering can also be described by ionization of electrons in atomic orbits. Therefore, the electron form factor is also depicted by ionization form factors $|f_{nl}(p_e,q)|^2$~\cite{Bunge1993RHF}. As pointed out in Ref.~\cite{An:2025cik}, since the germanium target is a single crystal, the scatterings of electrons in the same state around different atoms interfere with each other, leading to coherent scattering with discrete momentum transfer corresponding to reciprocal lattice vectors $\mathbf{G}$, which resembles the Bragg scattering of X-rays in crystals. After considering the effect of Bragg scattering in the calculation of expected event rates, the traditional results should be corrected, especially for $m_\chi\lesssim 0.05\,\text{MeV}$, since $\sqrt{2m_\chi\times 160\,\text{eV}}\lesssim3\,\text{keV}\sim\text{the lattice basis of }\mathbf{G}$. Meanwhile, for valence shell scatterings, the form factors are derived from Bloch state wave functions in the crystal, which are numerically computed by \textsc{Quantum ESPRESSO}~\cite{Giannozzi2017AdvancedCF}, with lattice parameters taken from Ref.~\cite{germanium}. The data and background model are taken from Ref.~\cite{CDEX:2018lau}. CDEX-10 has a recoil energy detection threshold of 160 eV, so we consider event counts corresponding to exposure time $T$ between 160 eV and 1500 eV:
\begin{equation}
    N=T\int_{160\,\text{eV}}^{1500\,\text{eV}}dE_e\left(\frac{dR_{\text{in}}}{dE_e}+\frac{dR_{\text{val}}}{dE_e}\right).
\end{equation}
With the measured event counts $S$ and background model counts $B$, we can analyze the constraints with a simplified statistical model. The signal event counts $S$ are treated as Poisson random variables, with expectation values $\mu=N+B$ and variances $\sigma^2=N+B$. If $N+B\gg 1$, they can be well approximated by Gaussian random variables. In this way, the likelihood ratio statistic can be constructed as
\begin{equation}
    -2\ln\lambda=-\ln\left(\frac{S}{N+B}\right)+\frac{(S-N-B)^2}{N+B}.
\end{equation}
For a sufficiently large sample, this statistic approaches a $\chi^2(1)$-distributed random variable. If $-2\ln\lambda>3.841$, the hypothesis of the DM model is rejected. 

\begin{figure}
    \centering
    \subfigure{
        \includegraphics[width=0.47\linewidth]{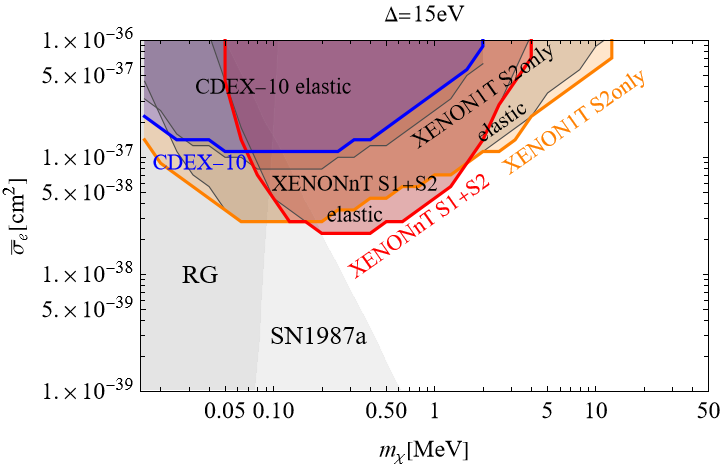}
    }
    \subfigure{
        \includegraphics[width=0.47\linewidth]{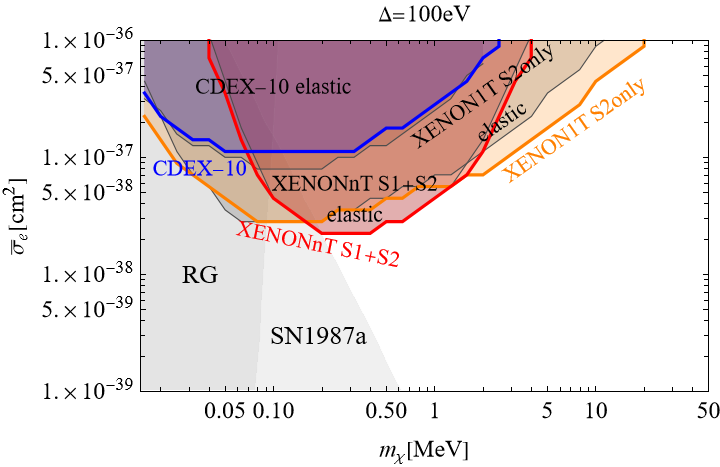}
    }
    \subfigure{
        \includegraphics[width=0.47\linewidth]{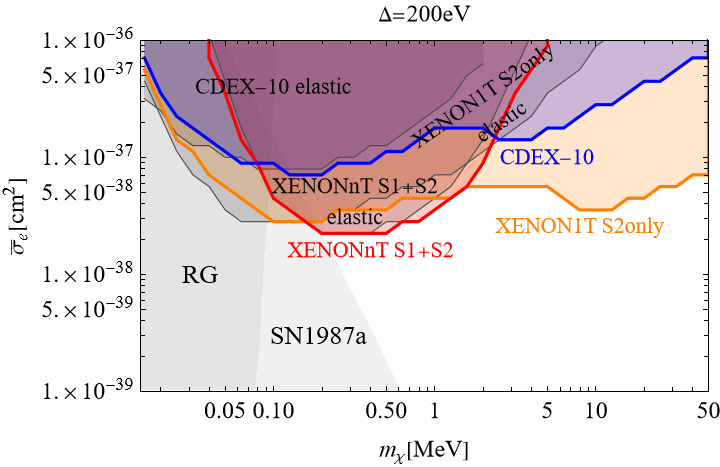}
    }
    \subfigure{
        \includegraphics[width=0.47\linewidth]{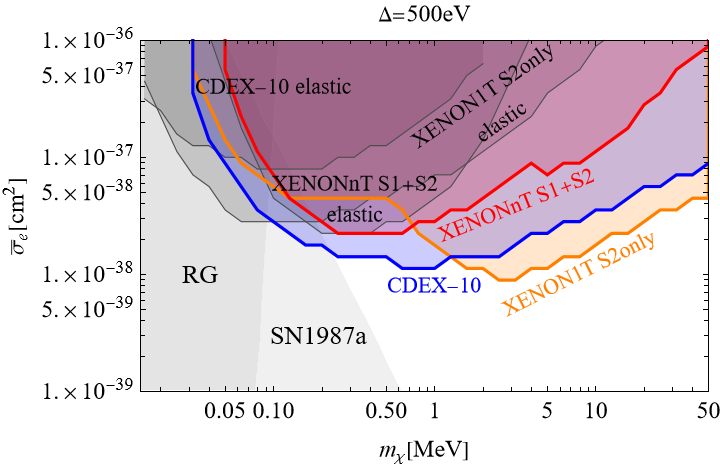}
    }
    \subfigure{
        \includegraphics[width=0.47\linewidth]{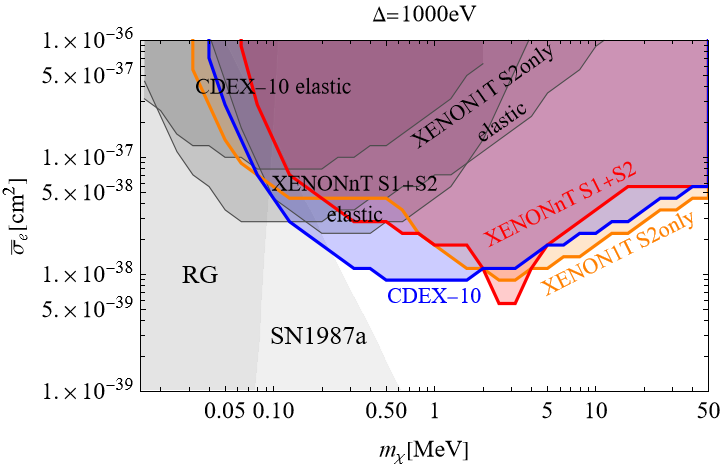}
    }
    \subfigure{
        \includegraphics[width=0.47\linewidth]{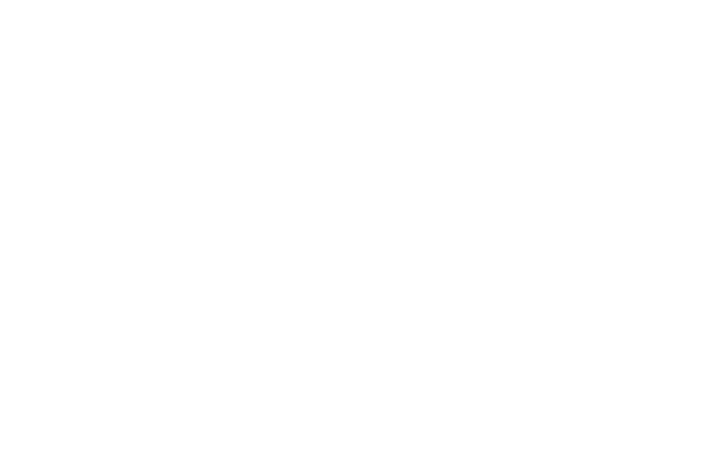}
    }
    \caption{This group of graphs illustrates constraints derived from XENONnT S1+S2, XENON1T S2only and CDEX-10 data for different mass splittings $\Delta$. Elastic constraints and astrophysical constraints are also shown for comparison. RG stands for red giant cooling constraints from Ref.~\cite{Chang:2018rso}. SN1978a stands for supernova SN1987a constraints from Ref.~\cite{Chang:2019xva}, assuming a dark photon-mediated model with dark coupling $\alpha_D=5$ and dark photon mass $m_V=3m_\chi$.}
    \label{fig:scanplots}
\end{figure}

\begin{figure}
    \centering
    \includegraphics[width=0.7\linewidth]{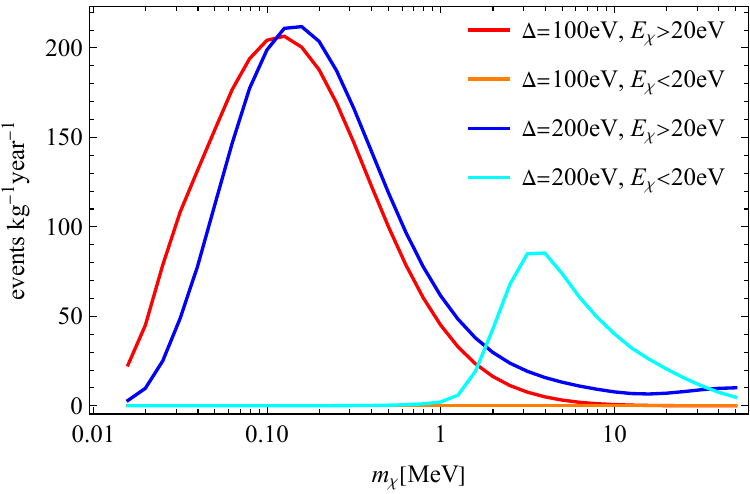}
    \caption{The expected event numbers in CDEX-10 are divided into contributions from reflected DM with $E_\chi<20\,\text{eV}$ and $E_\chi>20\,\text{eV}$. The reference cross section is selected to be $\bar{\sigma}_{\text{tot}}=1.58489\times 10^{-37}\,\text{cm}^2$ for all points in this graph. At $\Delta=100\,\text{eV}$, the $E_\chi<20\,\text{eV}$ part does not have enough energy to prompt the recoil energy of electron to exceed the $160\,\text{eV}$ electron recoil threshold. At $\Delta=200\,\text{eV}$, with sufficiently large DM mass, the transfer of internal energy from mass splitting into electron recoil energy is efficient enough such that the contribution of $E_\chi<20\,\text{eV}$ becomes the dominant contribution. Therefore, we can see a second peak around $m_\chi\sim 5m_e$.}
    \label{fig:divide_plot}
\end{figure}

The exclusion region is shown in Fig.~\ref{fig:scanplots}. We show the result for five benchmark mass splittings, $\Delta=15,100,200,500,1000\,\text{eV}$. For $\Delta\ll 15\,\text{eV}$, both the solar temperature and the detection threshold are much larger than the mass splitting; hence both solar reflection and detection are only slightly affected by the mass splitting. In this situation, the signal is almost the same as in the elastic case. For $\Delta\gg 1\,\text{keV}$, the mass splitting is larger than the core temperature of the Sun, so there are not enough high-energy electrons in the Sun to exceed the mass splitting. Therefore, solar reflection is greatly suppressed. From Fig.~\ref{fig:scanplots} we can see that the detection sensitivity of CDEX-10 is greatly augmented for $\Delta\geq 200\,\text{eV}$ because, at this point, for excited DM scattering in the detector, the mass splitting provides enough energy to exceed the 160 eV electron recoil threshold. When $\Delta=200\,\text{eV}$, apart from the maximum sensitivity around $m_\chi\approx m_e$ where the energy transfer of DM-electron scatterings in the Sun is the most efficient, there is also a weak maximum around $m_\chi\approx 5m_e$. As mentioned above, when $m_\chi>m_e$ the signal is dominated by mass splitting release. At this mass, the population of DM that does not gain much energy from electron scatterings or is only scattered by ions in the Sun reaches kinetic energy $\sim 20\,\text{eV}$, as shown in Fig.~\ref{fig:divide_plot}. Second, although the total solar reflected flux is suppressed, the mass splitting allows SRDM with energy $\sim 20\,\text{eV}$ to reach the threshold (as shown in Fig.~\ref{fig:detector_scat_a2}), compensating for the suppression of overall flux. Meanwhile, the XENONnT S1+S2 sensitivity also receives a great enhancement when $\Delta\geq500\,\text{eV}$. For $\Delta=1000\,\text{eV}$, we also see a maximum sensitivity around $m_\chi\approx 5m_e$. This can also be explained by the fact that the $1\,\text{keV}$ mass splitting allows the low-kinetic-energy SRDM to reach the $\sim 1\,\text{keV}$ threshold of XENONnT S1+S2 signals.

\section{Summary}
For light DM with a mass similar to that of the electron, a keV energy tail in the spectrum is an inevitable consequence of solar reflection. This tail flux increases the electron recoil energy and helps us to impose stringent constraints on the DM-electron contact interaction cross section of light DM models with $m_\chi\sim1\,\text{MeV}$ down to $\bar{\sigma}_e\sim 10^{-38}\,\text{cm}^2$ with the data of direct-detection experiments. In this paper, we extend this idea by introducing a general inelastic DM model with two mass states. The scattering process in the Sun not only increases the kinetic energy but also transforms the DM into an excited state. The DM in the excited state can then release additional energy in the scattering with electrons from the target material. As a consequence, the energy deposit is immediately increased. Because of the conservation of energy and momentum in an inelastic scattering process, the additional energy release is carried away mostly by the particle with smaller mass. Therefore, for DM with $m_\chi<m_e$, the additional energy release from the excited state cannot be converted into electron recoil energy. Consequently, the mass splitting does not improve sensitivity. However, for DM with $m_\chi>m_e$, the energy release can be converted to the electron recoil energy efficiently, resulting in a great improvement in sensitivity. In this paper, the scattering rates between solar electrons (ions) and inelastic DM are calculated. With the expressions of the scattering rates, the solar-reflected spectra of DM in both the ground and excited state are generated by a Monte Carlo simulation program. In the end, the constraints on inelastic DM models with different mass splittings are calculated using the data from the CDEX-10 and XENONnT experiments. In the case of solar reflected elastic DM, the sensitivity of DM with $m_\chi\sim 2$ to $10\,\text{eV}$ is weak due to the inefficiency of energy transfer between solar electrons and DM particles. However, with the help of the additional energy release from the solar excited inelastic DM, the exclusion bound can be imposed down to $\bar{\sigma}_e\sim10^{-38}\text{cm}^2$.

\appendix
\section{Derivation of the Scattering Amplitude}
In Section 3, we write down an inelastic DM amplitude without model details to maintain generality. Inelastic DM can be derived from many different models~\cite{Lu:2023cet,Kang:2021oes}, here we only consider an effective Lagrangian involving two real scalars $\chi_1$ and $\chi_2$:
\begin{equation}
    \mathcal{L}=\frac{1}{2}\partial_\mu\chi_1\partial^\mu\chi_1+\frac{1}{2}\partial_\mu\chi_2\partial^\mu\chi_2-\frac{1}{2}m_{\chi_1}^2\chi_1^2-\frac{1}{2}m_{\chi_2}^2\chi_2^2-g\chi_1\chi_2\bar{\psi}\psi,
\end{equation}
where $\psi$ is the electron field. The non-relativistic amplitude of the up-scattering process $\chi_1(k_1)+e^-(p_1)\rightarrow\chi_2(k_2)+e^-(p_2)$ can be easily derived:
\begin{equation}
    \sum_{\text{spin}}|M|^2=2g^2(p_2\cdot p_1+m_e^2)\approx4 g^2 m_e^2.
\end{equation}
If $\Delta\ll m_\chi$, the total cross section is
\begin{equation}
    \bar{\sigma}_{\text{tot}}=\frac{1}{4m_\chi m_e |v_\chi-v_e|}\int\mathrm{d}\Omega\frac{p_{\text{CM}}}{(2\pi)^2 4(m_\chi+m_e)}\sum_{\text{spin}}|M|^2=\frac{4g^2m_e^2}{16\pi(m_\chi+m_e)^2},\label{barsigmatot}
\end{equation}
where $p_{\text{CM}}$ is the momentum of one incoming particle in the center of mass frame. If $\Delta\ll m_\chi$, the down-scattering cross section is the same as in Eq.~\eqref{barsigmatot}. Therefore, the amplitude can be rewritten as
\begin{equation}
    \sum_{\text{spin}}|\mathcal{M}|^2=16\pi(m_\chi+m_e)^2\bar{\sigma}_{\text{tot}}.
\end{equation}

For couplings with other particles, such as protons or Helium nuclei, the calculation is similar, but the mass, coupling and spin sum are changed according to the effective Lagrangian.

\acknowledgments

This work is supported in part by the National Key R\&D Program of China under Grant Nos. 2023YFA1607104 and 2021YFC2203100, and the National Science Foundation of China under Grant Nos. 12475107 and 12525506.

%\paragraph{Note added.} This is also a good position for notes added after the paper has been written.

% Bibliography

%% [A] Recommended: using JHEP.bst file
\bibliographystyle{JHEP}
\bibliography{biblio.bib}

%% or
%% [B] Manual formatting (see below)
%% (i) We suggest to always provide author, title and journal data or doi:
%% in short all the informations that clearly identify a document.
%% (ii) please avoid comments such as "For a review'', "For some examples",
%% "and references therein" or move them in the text. In general, please leave only references in the bibliography and move all
%% accessory text in footnotes.
%% (iii) Also, please have only one work for each \bibitem.

\end{document}